\documentclass[sigconf]{acmart}

\usepackage{ctable}
\usepackage{multirow}
\usepackage{balance}
\usepackage{booktabs}
\usepackage{hyphsubst}
\usepackage{graphicx}
\usepackage{amsmath}
\usepackage{url}
\usepackage{breakurl}
\usepackage{hhline}
\usepackage[caption = false]{subfig}

\newcounter{lizcounter}
\DeclareRobustCommand{\liz}[1]{\textbf{/* #1 (liz) */}\stepcounter{lizcounter}\typeout{LaTeX Warning: liz comment \thelizcounter: #1 (line \the\inputlineno)}}

\newcounter{findingscounter}

\newcommand{\tr}{\emph{TagRec}}

\newcommand{\para}[1]{\vspace{2mm}\noindent\textbf{#1}}

\begin{document}

\title{The TagRec Framework as a Toolkit for the Development of Tag-Based Recommender Systems}

\author{Dominik Kowald}
\affiliation{%
  \institution{Know-Center \& Graz University of Technology}
}
\email{dkowald@know-center.at}
\author{Simone Kopeinik}
\affiliation{%
  \institution{Graz University of\\Technology}
}
\email{simone.kopeinik@tugraz.at}
\author{Elisabeth Lex}
\affiliation{%
  \institution{Graz University of\\Technology}
}
\email{elisabeth.lex@tugraz.at}

\begin{abstract}
Recommender systems have become important tools to support users in identifying relevant content in an overloaded information space. To ease the development of recommender systems, a number of recommender frameworks have been proposed that serve a wide range of application domains. Our \tr{} framework is one of the few examples of an open-source framework tailored towards developing and evaluating \emph{tag-based recommender systems}. In this paper, we present the current, updated state of \tr{}, and we summarize and reflect on four use cases that have been implemented with \tr{}: (i) tag recommendations, (ii) resource recommendations, (iii) recommendation evaluation, and (iv) hashtag recommendations. To date, \tr{} served the development and/or evaluation process of tag-based recommender systems in two large scale European research projects, which have been described in 17 research papers. Thus, we believe that this work is of interest for both researchers and practitioners of tag-based recommender systems.
\end{abstract}


\maketitle

\section{Introduction}
Recommender systems aim to predict the probability that a specific user will like a specific resource. Therefore, recommender systems utilize the past user behavior (e.g., resources previously consumed by this user) in order to generate a personalized list of potentially relevant resources \cite{ricci2011introduction}. Popular application domains of recommender systems include online marketplaces (e.g., Amazon and Zalando), movie and music streaming services (e.g., Netflix and Spotify), job portals (e.g., LinkedIn and Xing), and social tagging systems (e.g., BibSonomy and CiteULike).

Social tagging systems bear particularly great potential for recommender systems as, by nature, they produce a vast amount of user-generated resource-annotations (i.e., tags). Thus, possible use cases of these tag-based recommender systems include the suggestion of resources to extend a user's set of bookmarks \cite{de2010query,zanardi2008social} and the suggestion of tags to assist in the annotation of these bookmarks. The latter one is known as the field of tag recommendations \cite{jaschke2007tag}.

Over the past years, various recommendation frameworks and libraries have been developed in order to support the development and evaluation of recommender systems (see Section \ref{s:relwork}). While these frameworks cover a wide range of application domains, to the best of our knowledge, an open-source recommendation framework to design and evaluate tag-based recommender systems was still lacking. Therefore, in 2014, we have started developing \tr{}, a standardized tag recommender benchmarking framework \cite{kowald2014tagrec}. In the initial development phase of the framework, we mainly focused on evaluating tag recommendation algorithms. In 2015, the framework was extended by including resource recommendation algorithms that are based on social tagging data \cite{trattner2015tagrec}.

The aim of this paper, however, is to present the current, updated state of \tr{}. This includes the extension of the framework for (i) the analysis of tag reuse practices \cite{kowald2016influence}, (ii) the evaluation of tag recommendations in real-world folksonomy and Technology Enhanced Learning settings \cite{kowald2015evaluating,kopeinik2016algorithms}, and (iii) hashtag recommendations in Twitter \cite{www2017}.

Apart from that, we provide an updated framework description (see Section \ref{s:framework}) as well as a summary of use cases in the field of recommender research that have been completed using \tr{} (see Section \ref{s:use}). Research areas encompass tag recommendations, resource recommendations, recommendation evaluation and hashtag recommendations. To date, \tr{} has served the recommender development and/or evaluation processes in two large-scale European research projects, which have been published in 17 research papers. We conclude the paper with a discussion on future work and potential improvements of the framework (see Section \ref{s:conc}).

We believe that our work contributes to the rich portfolio of technical frameworks in the area of recommender systems. Furthermore, this paper presents an overview of use cases which can be realized with \tr{}, and should be of interest for both researchers and developers of tag-based recommender systems. 

\section{TagRec} \label{s:framework}
\tr{} is a Java-based recommendation framework for tag-based information retrieval settings. It is open-source software and freely available via our Github repository\footnote{\url{https://github.com/learning-layers/TagRec}}. The Github page also contains a detailed technical description on the usage of the framework.

\begin{figure}[t]
   \centering 
   \includegraphics[width=0.48\textwidth]{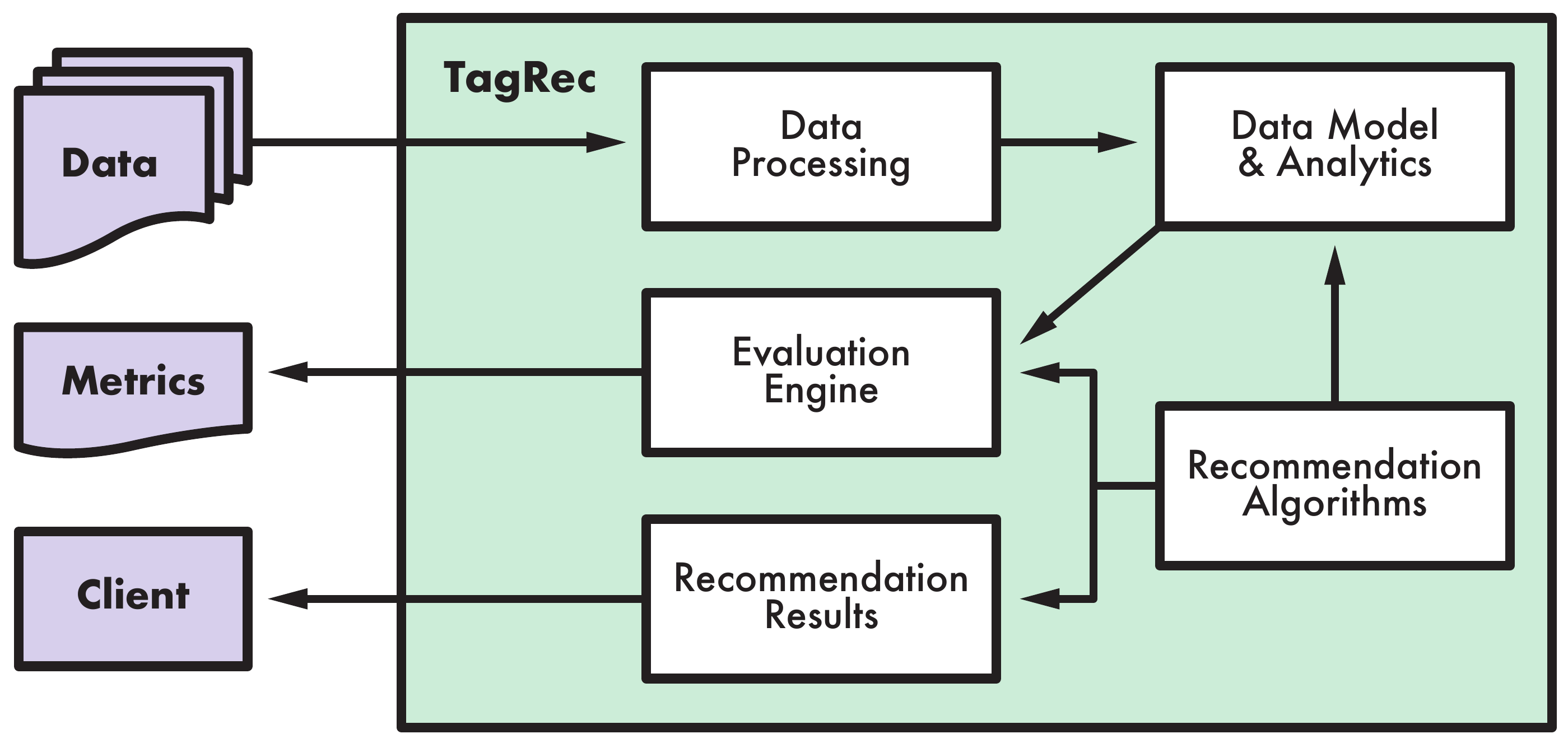} 
   \caption{System architecture of \tr{}. Here, the data processing component processes data sources in order to create a data model and data analytics. Then, this data model is used by recommendation algorithms to create recommendation results that are either forwarded to an evaluation engine or to a client application.
	\vspace{-3mm}
	}
	 \label{fig:tr}
\end{figure}

Figure \ref{fig:tr} illustrates \tr{}'s system architecture. The framework consists of (i) a data processing component, which processes data sources, (ii) a data model and analytics component, which enables access to the processed data, (iii) recommendation algorithms, which calculate recommendations, (iv) an evaluation engine, which evaluates the algorithms, and (v) recommendation results, which can be passed to a client application. The mentioned components are described in more detail in the remainder of this section. Apart from that, we describe practical aspects of the framework that should be helpful when implementing and/or evaluating a recommendation algorithm. Finally, Table \ref{tab:stats} provides an overview of the supported datasets, recommendation algorithms and evaluation metrics. 

\para{Data Processing.}
The data processing component is responsible for parsing and processing external data sources. Currently supported datasets are listed in Table \ref{tab:stats}. These datasets serve a wide range of application domains such as social bookmarking systems, learning environments, microblogging tools and music/movie sharing portals. The set of datasets can easily be extended by implementing custom data pre-processing strategies.

Furthermore, this component supports various data enrichment and transformation methods such as $p$-core pruning \cite{doerfel2013analysis}, topic modeling \cite{krestel2009latent}, training/test set splitting \cite{kowald2015evaluating} and data conversion into related formats (e.g., for MyMediaLite \cite{gantner2011mymedialite}).

\begin{table}[t]
  \setlength{\tabcolsep}{3.5pt}
  \centering
    \begin{tabular}{l|l}
    \specialrule{.2em}{.1em}{.1em}
    \textbf{Dataset}	  													& \textbf{Description}				\\\hline 
			Flickr		 & 	Image sharing \cite{kowald2015evaluating}								\\      
			CiteULike	 & 	Scientific references \cite{kowald2015evaluating}							\\
			BibSonomy	 & 	Publication sharing \cite{kowald2015evaluating}							\\						
			Delicious	 & 	Social bookmarking \cite{kowald2015evaluating}								\\
			LastFM		 & 	Music sharing \cite{kowald2015evaluating}								\\
			MovieLens	 & 	Movie rating \cite{kowald2015evaluating}							\\
			Twitter		 & 	Microblogging	\cite{www2017}						\\
			TravelWell & 	Learning resource exchange \cite{kopeinik2016algorithms}									\\
			Aposdle    & 	Work-integrated learning \cite{kopeinik2016algorithms}					\\
			MACE			 & 	Informal learning \cite{kopeinik2016algorithms}						\\
			KDD15			 &	KDD 2015 cup \cite{kopeinik2016algorithms}						\\
   \specialrule{.2em}{.1em}{.1em} 
   \textbf{Algorithm}	  													& \textbf{Description}				\\\hline 
      MostPopular & 		Frequency-based	\cite{kowald2016influence}					\\
			CF & Collaborative Filtering \cite{marinho2008collaborative}							\\						
			FolkRank / APR & 	Graph-based	\cite{jaschke2007tag}							\\
			FM / PITF & 	Factorization Machines \cite{rendle2010pairwise}							\\
			LDA & 		Topic modeling \cite{krestel2009latent}						\\
			MostRecent / GIRP & 	Time-based \cite{kowald2016influence}								\\			
			3Layers & 	Human categorization theory \cite{seitlinger2013recommending,kowald2015forgetting,kowald2015modeling}					\\
			BLL / BLL$_{AC}$ & 	Human memory theory \cite{kowald2014long,kowald2015refining,trattner2016modeling,www2017}							\\
			CIRTT & Tag- and time-based	\cite{ht2014}				\\
			SUSTAIN & Human category learning	\cite{seitlinger2015attention,kopeinik2017improving}						\\
			SimRank & Content-based \cite{zangerle2011recommending} \\
			BLL$_{I,S,C}$ & Temporal hashtag patterns \cite{www2017} \\
   \specialrule{.2em}{.1em}{.1em} 
   \textbf{Metric}	  													& \textbf{Description}				\\\hline 
      Recall & 		Accuracy \cite{kowald2015evaluating}						\\
			Precision & Accuracy \cite{kowald2015evaluating}								\\						
			F1-score & 	Accuracy \cite{kowald2015evaluating}								\\
			MRR & 	Ranking	\cite{kowald2015evaluating}							\\
			MAP & 	Accuracy \& ranking \cite{kowald2015evaluating}								\\
			nDCG & 		Accuracy \& ranking \cite{kowald2015evaluating}							\\
			AILD & 	Diversity	\cite{kowald2015evaluating}						\\
			AIP & 	Novelty	\cite{kowald2015evaluating}							\\
			Runtime & Computational costs \cite{kowald2015evaluating}							\\
			Memory & 	Computational costs \cite{kowald2015evaluating}								\\			
		\specialrule{.2em}{.1em}{.1em}								
    \end{tabular}
		\caption{Datasets, recommendation algorithms and evaluation metrics supported by \tr{}. A complete list of the features is provided on the framework's Github page.
		\vspace{-7mm}
		}
		\label{tab:stats}
\end{table}

\para{Data Model and Analytics.}
The data model is created based on described data processing steps and provides an object-oriented representation of the data in order to ease the implementation process of a novel recommendation algorithm. Thus, it enables easy access to the entities in the datasets via powerful query functionality (e.g., get the set of tags a user has used in the past). Furthermore, the data model of \tr{} is connected to Apache Solr\footnote{\url{http://lucene.apache.org/solr/}} and thus, enables fast access to content-based data of entities.

Another role of this component is the provision of basic data analytics functionality to get a better understanding of the dataset characteristics. For example, dataset statistics, such as the total number of distinct tags or the average number of bookmarks per user, can be retrieved.

\para{Recommendation Algorithms.}
\tr{} contains a wide range of recommendation algorithms (see Table \ref{tab:stats}). As later described in Section \ref{s:use}, algorithms for tag recommendations, resource recommendations and hashtag recommendations are provided. These algorithms can be used as baseline approaches for a newly implemented algorithm. The complete list of all variants of these algorithms is provided on the \tr{}'s Github page.

A key contributions of \tr{} is that next to well-established approaches, such as Collaborative Filtering, MostPopular and Factorization machines, it also encompasses approaches based upon cognitive models of information retrieval, human memory theory and category learning. In \cite{kowald2015evaluating}, it has been shown that these cognitive-inspired approaches achieve high prediction accuracy estimates in comparison to classic recommendation algorithms. Besides, the algorithms have demonstrated their suitability for sparse datasets such and narrow folksonomies.

\para{Evaluation Engine.}
The evaluation engine quantifies the quality of implemented recommendation strategies by applying a rich set of evaluation metrics as listed in Table \ref{tab:stats}. One drawback of most recommendation evaluation frameworks is their focus on accuracy and ranking estimates, which restricts the evaluation to the performance of recommender systems \cite{bordino2013penguins}.

To fill this gap, \tr{} supports a variety of evaluation metrics to also offer indicators for diversity, novelty, runtime performance and memory consumption of algorithms. For evaluating an algorithm, \tr{} has to be provided with three parameters, where the first one specifies the algorithm, the second one specifies the dataset directory and the third one specifies the file name of the dataset sample. For example, \textit{java \-jar tagrec.jar cf bib bib\_sample} runs Collaborative Filtering on a sample of the BibSonomy dataset. The calculated metrics are then either written to a ``metrics'' file or printed to the console. 

\begin{figure}[t]
   \centering 
   \includegraphics[width=0.48\textwidth]{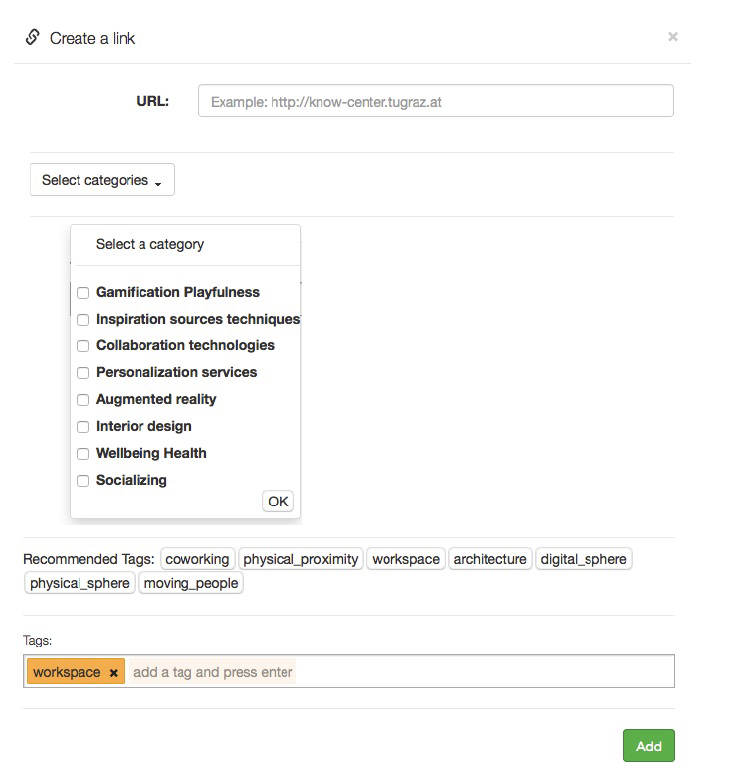} 
   \caption{Screenshot of a user interface for the online evaluation of tag recommendations using the \tr{} framework.
		\vspace{-3mm}	
}
	 \label{fig:kb}
\end{figure}

\para{Recommendation Results.}
As indicated in Figure \ref{fig:tr}, the algorithms' recommendation results can be either forwarded to the evaluation engine to retrieve evaluation metrics or to a client application for further processing (e.g., visualization). The KnowBrain tool \cite{dennerlein2015knowbrain} is an example of such a client application. It is an open source social bookmarking tool, which has been extended to cater the requirements of tag recommender evaluations in online settings.

A screenshot of KnowBrain's graphical user interface is shown in Figure \ref{fig:kb}. It enables the bookmarking of Web links and their annotations by (i) selecting from a pre-defined set of categories, and by (ii) assigning a variable number of tags. The user's tagging process is supported by a list of recommended tags that are selected based on algorithms of the \tr{} framework. 
The elicitation of categories allows for semantic context-based recommendation algorithms such as 3Layers \cite{seitlinger2013recommending}. Furthermore, the comparison of actually used tags with recommended tags gives insights into the online performance (i.e., user acceptance) of recommendation strategies. 

\section{Use Cases} \label{s:use}
In this section, we describe use cases that have been implemented using the \tr{} framework. To date, \tr{} supported the recommender development and/or evaluation processes in two large-scale European research projects. Results have been published in 17 research papers (see Table \ref{tab:usecases}).

\subsection{Tag Recommendations}
Tag recommendation systems assist users in finding descriptive tags to annotate resources. In other words, given a specific user and a specific resource, a tag recommendation algorithm predicts a set of tags a user is likely to apply in annotating the resource \cite{jaschke2007tag}. 

Within this context, \tr{} was used for the creation of (i) cognitive-inspired algorithms, and (ii) the evaluation of approaches suitable for formal and informal learning settings.

\begin{table}[t]
  \setlength{\tabcolsep}{4.0pt}
  \centering
    \begin{tabular}{l|l}
    \specialrule{.2em}{.1em}{.1em}
    \textbf{Tag recommendations}	  			& \textbf{Research papers}				\\\hline 
			Model of human categorization		 & 	\cite{seitlinger2013recommending,kowald2015forgetting,kowald2015modeling}							\\      
			Activation processes in human memory	 & 	\cite{kowald2016influence,kowald2014long,kowald2015refining,trattner2016modeling}							\\
			Informal learning settings & \cite{dennerlein2015social,dennerlein2015knowbrain,dennerlein2014making} \\
   \specialrule{.2em}{.1em}{.1em} 
    \textbf{Resource recommendations}	  			& \textbf{Research papers}				\\\hline 
      Attention-interpretation dynamics & \cite{seitlinger2015attention,kopeinik2017improving} \\
			Tag and time information & 		\cite{ht2014,larrain2015good}					\\
   \specialrule{.2em}{.1em}{.1em} 
    \textbf{Recommendation evaluation}	  			& \textbf{Research papers}				\\\hline 
      Real-world folksonomies & 		\cite{kowald2015evaluating}						\\
			Technology enhanced learning settings & \cite{kopeinik2016algorithms}						\\						
   \specialrule{.2em}{.1em}{.1em} 
    \textbf{Hashtag recommendations}	  			& \textbf{Research papers}				\\\hline 
      Temporal effects on hashtag reuse & 	\cite{www2017}						\\		
		\specialrule{.2em}{.1em}{.1em}								
    \end{tabular}
		\caption{Use cases realized with \tr{}. To date, \tr{} supported the recommender development and/or evaluation processes described in 17 research papers.
		\vspace{-5mm}
}
		\label{tab:usecases}
\end{table}

\para{Tag Recommendations Using a Model of Human Categorization.}
In \cite{seitlinger2013recommending,kowald2015forgetting,kowald2015modeling}, the authors introduced a tag recommendation algorithm based on the human categorization models ALCOVE \cite{kruschke1992alcove} and MINERVA2 \cite{hintzman1984minerva}. This algorithm is called \emph{3Layers} and simulates categorization processes in human memory. Therefore, the categories assigned to a given resource, which a user is going to annotate, are matched against already annotated resources of this user. Based on this matchmaking process, a set of tags associated with semantically related resources is recommended.

Since \tr{} enables to link a list of categories to a resource, it supported the development of \emph{3Layers} by providing functions for analyzing and deriving category information of resources (e.g., via LDA topic modeling \cite{krestel2009latent}).

\para{Utilizing Activation Processes in Human Memory.}
Activation processes in human memory describe the general and context-dependent usefulness of information. It was shown that these processes (especially usage frequency, recency and semantic context) greatly influence the reuse probability of tags \cite{kowald2016influence}. Based on this, a set of time-aware tag recommendation approaches (see \cite{kowald2014long,kowald2015refining,trattner2016modeling}) was developed that utilize the activation equation of the cognitive architecture ACT-R \cite{anderson2004integrated}.

Therefore, \tr{} was used to analyze the timestamps of tag assignments and to calculate tag co-occurrences for reflecting the semantic context of social tagging. Furthermore, \tr{} enabled the hybrid combination of the components of the model (e.g., combining time-aware and context-aware recommendations).

\para{Tag Recommendations in Informal Learning Settings.}
In the course of the European-funded project Learning Layers\footnote{\url{http://learning-layers.eu/}}, which aims at supporting informal learning at the workplace, tag recommendations were used to support the individual user in finding descriptive tags and the collective in consolidating a shared tag vocabulary. These tag recommendations were used in two tools: (i) the Dropbox-like environment KnowBrain \cite{dennerlein2015knowbrain}, and (ii) the Sensemaking interface Bits \& Pieces \cite{dennerlein2014making}.

To achieve this, \tr{} was integrated as a tag recommendation library into the Social Semantic Server \cite{dennerlein2015social}, which was used as the technical back-end for KnowBrain and Bits \& Pieces. This shows that \tr{} cannot only be used as a standalone tool but also as a programming library (or toolkit) to include recommendation functionality in existing software. A similar approach will be followed in another European-funded project called AFEL\footnote{\url{http://afel-project.eu/}}, which engages in design and development of analytics for everyday learning.

\subsection{Resource Recommendations}
Resource recommender systems suggest potentially relevant web items (e.g., movies, books, learning resources, URLs, etc.) to users. Most of these recommender systems are based on Collaborative Filtering (CF) techniques, which aim to calculate similarities between users to suggest the most suitable web resources to them \cite{sarwar2001item}. \tr{} was applied to support and improve the development of CF approaches in tag-based online environments.

\para{Mimicking Attention-Interpretation Dynamics.}
Seitlinger et al. \cite{seitlinger2015attention} introduced the first version of a CF-based recommendation approach that takes into consideration non-linear user-artifact dynamics, modeled by means of SUSTAIN. SUSTAIN (\textit{Supervised and Unsupervised STratified Adaptive Incremental Network}) is a flexible network model of human category learning that is thoroughly discussed in \cite{love2004sustain}. It assumes that learning is a dynamic process that takes place through the encounter of new examples (e.g., Web resources). Throughout the learning trajectory, categories emerge and learners' attention foci shift. In \cite{kopeinik2017improving}, an advanced, adapted version of the initial approach was presented and analyzed in detail. The resulting approach $SUSTAIN+CF$, firstly applies CF to calculate the most suitable resources for a user, and secondly re-ranks this list depending on a user's category learning model (i.e., SUSTAIN's user model).

The algorithm has been implemented and developed within the \tr{} framework. Features of the framework allowed for continuous evaluation and analysis of single factors of the model and with it, the associated change of recommendation performance. With this data, it was possible to gain deeper insight into the algorithmic approach and its parameters and thus, to further adapt the model to the requirements of our application area.   

\para{Resource Recommendations using Tag and Time Information.}
In \cite{ht2014}, the \textit{Collaborative Item Ranking Using Tag and Time Information (CIRTT)} approach was presented. CIRTT uses Collaborative Filtering to identify a set of candidate resources and re-ranks these candidate resources by incorporating tag and time information. This is achieved via the Base-Level-Learning (BLL) equation, which is one component of ACT-R's activation equation \cite{anderson2004integrated}.

Since \tr{} contains a full implementation of the activation equation, it could be easily adapted for the task of resource recommendations as well. Apart from that, \tr{} was used to compare \textit{CIRTTT} to other related resource recommendation methods (e.g., \cite{zheng2011recommender}). Another study of the recency effects in Collaborative Filtering recommender systems was provided in \cite{larrain2015good}.

\subsection{Recommendation Evaluation}
One of the most challenging tasks in the area of recommender systems, is the reproducible evaluation of recommendation results \cite{herlocker2004evaluating}. \tr{} aims to support this process by providing standardized data processing methods, baseline algorithm implementations, evaluation protocols and metrics.

\para{Evaluating Tag Recommendations in Real-World Folksonomies.}
Because of the sparse nature of social tagging systems, most tag recommendation evaluation studies were conducted using $p$-core pruned datasets. This means that all users, resources and tags, which do not appear at least $p$ times in the dataset, are removed. This clearly does not reflect a real-world folksonomy setting as shown by \cite{doerfel2013analysis}.

To overcome this problem, \tr{} was used in \cite{kowald2015evaluating} to compare a rich set of tag recommendation algorithms using a wide range of evaluation metrics on six unfiltered social tagging datasets (i.e., Flickr, CiteULike, BibSonomy, Delicious, LastFM and MovieLens). The results showed that the efficacy of a recommendation algorithm greatly depends on the given dataset characteristics, and that cognitive-inspired approaches provide the most robust results, even in sparse data folksonomy settings.

\para{Comparing Recommendation Algorithms in Technology Enhanced Learning Settings.}
Kopeinik et al. \cite{kopeinik2016algorithms} is another example of using \tr{} for the evaluation of a variety of algorithms on different offline datasets. The paper focused on technology-enhanced formal and informal learning environments, where due to fast changing domains and characteristic group learning settings, data is typically sparse. The evaluation was divided in two settings, the performance of (i) resource recommendation strategies, and (ii) tag recommendation strategies. In both cases, the authors compared the recommendation accuracy of a number of computationally-inexpensive recommendation algorithms on six offline datasets retrieved from various educational settings (i.e., social bookmarking systems, social learning environments and massive open online courses). Investigated approaches are either state-of-the-art recommendation approaches, or strategies that have been explicitly suggested in the context of TEL systems.  

To address the goals of this study, the \tr{} framework already provided a wide range of required functionality such as the implemented data processing component, evaluation metrics and state-of-the-art algorithms. In the context of this research paper, it was further extended by a couple of algorithms that are considered particularly relevant to learning settings and by additional statistics, which were needed to interpret evaluation results properly. 

\subsection{Hashtag Recommendations}
Over the past years, hashtags have become very popular in systems such as Twitter, Instagram and Facebook. Similar to social tags, hashtags are freely-chosen keywords to categorize resources such as Twitter posts (i.e., tweets). One of the biggest advantages of hashtags is that they can be easily used by integrating them in the tweet text. Unsurprisingly, this has led to the development of hashtag recommendation algorithms that aim to support users in applying the most descriptive hashtags to their tweets \cite{zangerle2011recommending}.

\para{Temporal Effects on Hashtag Reuse.}
In \cite{www2017}, a time-dependent and cognitive-inspired hashtag recommendation approach was proposed. In this paper, temporal effects on hashtag reuse in Twitter have been analyzed with the help of \tr{} in order to design a hashtag recommendation approach, which utilizes the BLL equation of the cognitive architecture ACT-R \cite{anderson2004integrated}. Therefore, \tr{} was extended with functions to access Apache Solr (see Section \ref{s:framework}), which enables the content-based analysis of tweets using TF-IDF (see \cite{www2017}).

\section{Related Work} \label{s:relwork}
In recent years, a multitude of recommendation engines have been created and made available either for commercial use or as open-source systems. In \cite{recEngines}, a well-structured overview of such approaches is presented. The author differentiates between Software-as-a-Service (i.e., SaaS), non-SaaS, open-source, academic and benchmarking recommender systems. We consider open-source and benchmarking frameworks that implement recommendation strategies fitting academic purposes as most relevant to our work. 

A considerable contribution to this area is \emph{LibRec}\footnote{\url{http://wiki.librec.net/doku.php}}, a Java-based library that, so far, comprises around 70 resource recommendation algorithms and evaluation modules \cite{guo2015librec}. Another Java-based, open-source framework is \emph{RankSys}\footnote{\url{http://ranksys.org/}}, which focuses on the evaluation of ranking problems and supports the investigation of novelty as well as diversity for academic research \cite{castells2015novelty}, which is reflected in its design (e.g., data input interfaces work with a triple of user, item and features).

Other examples of open-source recommender software are \emph{MyMediaLite}\footnote{\url{http://www.mymedialite.net/}}, an item recommender library that focuses on rating and ranking predictions in collaborative filtering approaches \cite{gantner2011mymedialite}, \emph{CARSKit}\footnote{\url{https://github.com/irecsys/CARSKit}}, a recommendation library specifically designed for context-aware recommendations, and \emph{Tag Recommender}\footnote{\url{http://www.libfm.org/tagrec.html}}, a software component that implements Tensor Factorization models for personalized tag recommendations in C++ \cite{rendle2010pairwise}.  

However, to the best of our knowledge, an open-source recommender framework that implements a wide range of tag and resource recommendation algorithms (including a number of cognitive-inspired approaches) for the design and evaluation of personalized tag-based recommendation strategies was still missing.

\section{Conclusion and Future Work} \label{s:conc}
In this paper, we presented the \tr{} framework as a toolkit for the development and evaluation of tag-based recommender systems. \tr{} is open-source software written in Java and can be freely downloaded from Github. The framework consists of five components: (i) a data processing component, which processes data sources, (ii) a data model and analytics component, which enables access to the processed data, (iii) recommendation algorithms, which calculate recommendations, (iv) an evaluation engine, which evaluates the algorithms, and (v) recommendation results, which can be passed to client applications.

Apart from that, we summarized various use cases realized with \tr{} from the fields of tag recommendations, resource recommendations, recommendation evaluation and hashtag recommendations. To date, \tr{} supported the development and/or evaluation process described in 17 research papers. Specifically, our framework was used for the realization of recommendation algorithms based on models of cognitive science. In these papers, it was shown that the cognitive-inspired approaches provided the most robust results, even in sparse data folksonomy settings.

We believe that \tr{} extends the already rich portfolio of recommender frameworks with a toolkit that is specifically tailored to fit tag-based settings. Furthermore, the presentation of \tr{}'s use cases should be of interest for both researchers and developers of tag-based recommender systems.

\para{Limitations \& future work.} Currently, one limitation of \tr{} is that the data access is not standardized. Thus, social tagging data is accessed from folksonomy files, whereas resource-related metadata (e.g., tweet content) is accessed from Apache Solr.

Thus, our first plan for future work is to implement a mechanism that integrates all data into Apache Solr. Apart from that, we want to further work on the stability and code quality of the framework. For example, we want to enhance the build and dependency management of the software using Apache Maven\footnote{\url{https://maven.apache.org/}}.

\section{Acknowledgments}
This work was supported by the Know-Center Graz, the European-funded projects AFEL (GA: 687916) and Learning Layers (GA: 318209), and the Austrian Science Fund (FWF) project OMFix (Grant No P27709-G22). The Know-Center Graz is funded within the Austrian COMET Program - Competence Centers for Excellent Technologies - under the auspices of the Austrian Ministry of Transport, Innovation and Technology, the Austrian Ministry of Economics and Labor and by the State of Styria. COMET is managed by the Austrian Research Promotion Agency (FFG).

\balance

\end{document}